\documentclass[aps,pra,twocolumn,superscriptaddress,10pt,longbibliography]{revtex4-2}
\usepackage{amsmath,amssymb,bm}
\usepackage{graphicx}
\usepackage{epstopdf}
\usepackage{latexsym}
\usepackage[normalem]{ulem} 
\usepackage[caption=false]{subfig}
\usepackage[usenames,dvipsnames]{color}
\usepackage{hyperref}
\usepackage{mathtools}
\usepackage{natbib}
\usepackage{soul}
\usepackage{bbold}
\begin{document}

\newcommand{\Tr}[1]{\operatorname{Tr}( #1 )}
\newcommand{\ket}[1]{\lvert #1 \rangle}
\newcommand{\bra}[1]{\langle #1 \rvert}
\newcommand{\ketbra}[2]{\ket{#1}\bra{#2}}
\newcommand{\expt}[1]{\langle #1 \rangle}
\renewcommand{\mod}[1]{\lvert #1 \rvert}
\newcommand{\modsq}[1]{\mod{#1}^2}
\newcommand{\partialD}[2]{\frac{\partial #1}{\partial #2}}
\newcommand{\braket}[2]{\langle #1 | #2 \rangle}
\newcommand{\ns}{\mathcal{N}_{\mathrm{s}}}
\newcommand{\sinc}{\mathrm{sinc}}
\newcommand{\nb}{\mathcal{N}_{\mathrm{b}}}

\DeclarePairedDelimiter{\ceil}{\lceil}{\rceil}

\makeatletter
\def\maketitle{
\@author@finish
\title@column\titleblock@produce
\suppressfloats[t]}
\makeatother

\newif\ifincludeSupplement
\includeSupplementtrue

\newcommand{\mytitle}{Global Variational Quantum Circuits for Arbitrary Symmetric State Preparation}

\title{\mytitle}
\date{\today}

\newcommand{\affA}{Institute of Physics, University of Amsterdam, Science Park 904, 1098 XH Amsterdam, the Netherlands}
\newcommand{\affB}{QuSoft, Science Park 123, 1098 XG Amsterdam, the Netherlands}
\newcommand{\affC}{CWI, Science Park 904, 1098 XH Amsterdam, the 
Netherlands}
\newcommand{\affZ}{ARC Centre of Excellence for Engineered Quantum Systems, School of Mathematics and Physics, University of Queensland, St Lucia, QLD 4072, Australia}

\title{\mytitle}
\date{\today}

\author{Liam J. Bond}\email{L.J.Bond@uva.nl}
\affiliation{\affA}\affiliation{\affB}\affiliation{\affZ}
\author{Matthew J. Davis}\affiliation{\affZ}
\author{Ji\v{r}\'{i} Min\'{a}\v{r}}\affiliation{\affA}\affiliation{\affB}\affiliation{\affC}
\author{Rene Gerritsma}\affiliation{\affA}\affiliation{\affB}
\author{Gavin K. Brennen}
\affiliation{ARC Centre of Excellence for Engineered Quantum Systems, School of Mathematical and Physical Sciences, Macquarie University, Sydney, NSW 2109, Australia}
\author{Arghavan Safavi-Naini}
\affiliation{\affA}\affiliation{\affB}\affiliation{\affZ}

\begin{abstract}
Quantum states that are symmetric under particle exchange play a crucial role in fields such as quantum metrology and quantum error correction. We use a variational circuit composed of global one-axis twisting and global rotations to efficiently prepare arbitrary symmetric states, i.e. any superposition of Dicke states. The circuit does not require local addressability or ancilla qubits and thus can be readily implemented in a variety of experimental platforms including trapped-ion quantum simulators and cavity QED systems. We provide analytic and numerical evidence that any $N$-qubit symmetric state can be prepared in $2N/3$ steps. We demonstrate the utility of our protocol by preparing (i) metrologically useful $N$-qubit Dicke states of up to $N = 300$ qubits in $\mathcal{O}(1)$ gate steps with theoretical infidelities $1-\mathcal{F} < 10^{-3}$, (ii) the $N = 9$ Ruskai codewords in $P = 4$ gate steps with $1-\mathcal{F} < 10^{-4}$, and (iii) the $N = 13$ Gross codewords in $P = 7$ gate steps with $1-\mathcal{F} < 10^{-4}$. Focusing on trapped-ion platforms, for the $N = 9$ Ruskai and $N = 13$ Gross codewords we estimate that the protocol achieves fidelities $\gtrsim 95\%$ in the presence of typical experimental noise levels, thus providing a pathway to the preparation of a wide range of useful highly-entangled quantum states. 
\end{abstract}

\maketitle  

The operation of a quantum computer relies on our ability to prepare highly entangled quantum states, both as a step for quantum algorithms, and as codewords for quantum error correction (QEC). Despite recent remarkable experimental progress towards fault-tolerant quantum computing~\cite{bluvsteinLogicalQuantumProcessor2023,krinnerRealizingRepeatedQuantum2022,sivakRealtimeQuantumError2023}, the fragility of multi-partite entanglement to noise limits the fidelity of state preparation. Furthermore, the lack of single-site addressing in some experimental platforms has typically limited the variety of states that can be prepared to e.g. squeezed states. In light of these limitations, the choice of QEC scheme is important. Depending on its properties, certain QEC schemes may offer advantages both in terms of the types of errors that they can protect against~\cite{RevModPhys.87.307}, as well as their suitability of implementation to the particular hardware. One promising approach utilizes codewords which are symmetric states, i.e. superpositions of Dicke states~\cite{PhysRevA.90.062317,PhysRevLett.85.194,PhysRevLett.127.010504}. The Dicke states are also relevant for quantum metrology \cite{PhysRevA.85.022322}, including error-corrected quantum sensing \cite{ouyang2023quantum}, quantum storage \cite{PhysRevA.81.032317,PhysRevLett.97.170502} and quantum networking \cite{PhysRevLett.103.020503}. The need to overcome the intrinsic noise in NISQ devices has motivated the design of symmetric state preparation protocols that minimize the circuit depth. In particular, symmetric states have been prepared using one- and two-qubit gate circuits with depth $\mathcal{O}(N)$~\cite{9951196,9774323}, and specific Dicke states with constant depth using measurements and classical computation~\cite{buhrmanStatePreparationShallow2024}.

In this article, we demonstrate that \emph{arbitrary} superpositions of Dicke states can be engineered using linear depth circuits with only global control. Specifically, we utilize a variational quantum circuit consisting of only global one-axis twisting and global rotations, and thus do not demand local addressability nor ancilla qubits. We provide analytic and numerical evidence that the number of steps to synthesize an arbitrary $N$-qubit symmetric states scales as $2N/3$. Focusing on applications to QEC codewords, we show that only $P = 4$ and $P = 7$ steps are required to prepare QEC codewords of the $N = 9$ Ruskai \cite{PhysRevLett.85.194} and $N = 13$ Gross \cite{PhysRevLett.127.010504} codes with infidelity $1-\mathcal{F} < 10^{-4}$, respectively. We discuss the possible implementation in trapped ions, including a detailed analysis of expected sources of error. The limited resources required by our protocol make it immediately realizable on a broad range of experimental quantum simulation platforms.

\begin{figure*}
\includegraphics[width=\textwidth]{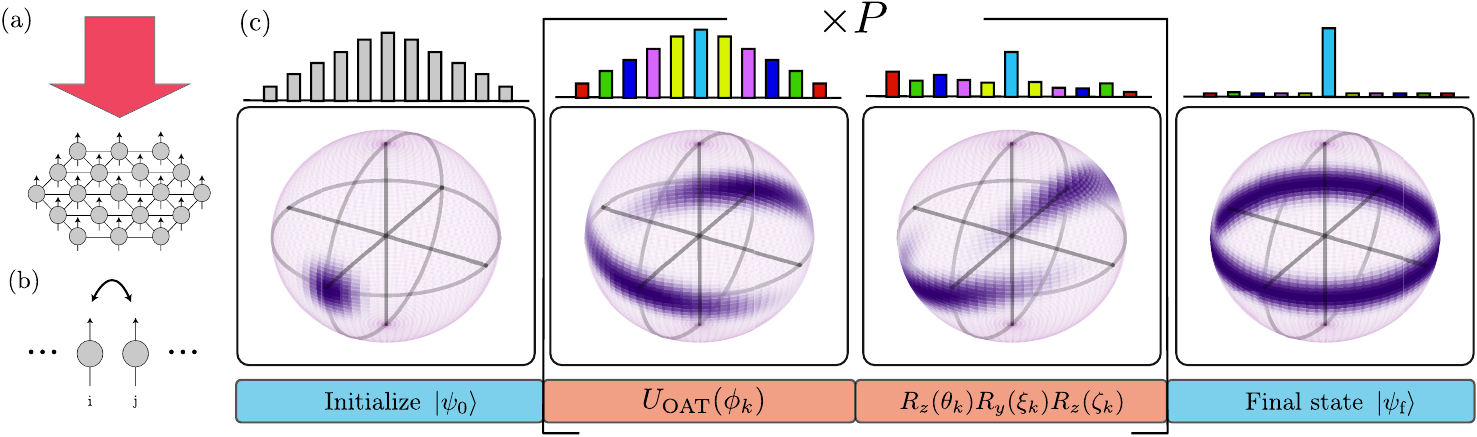}
\caption{
(a) The protocol Eq.~\eqref{eq:GateSequence} utilizes operations that act globally on a collection of $N$ qubits. (b) The Dicke states are symmetric, i.e. any two qubits $i$ and $j$ can be interchanged without changing the quantum state. (c) Schematic representation of the protocol. The system is initialized in a coherent spin state (CSS) $\ket{\psi_0} = \ket{\bar{\theta},\bar{\phi}}$. Next, one-axis twisting $U_\text{OAT}$ generates spin squeezing. For relatively small $\phi_k$ the CSS is spin squeezed; for larger $\phi_k$ the probability distribution separates across the Bloch sphere, as is the case sketched here. Next, the state is rotated about an arbitrary axis. These two operations are repeated $P$ times, with parameters numerically optimized to minimize the infidelity $1-\mathcal{F}$. Above the Bloch spheres, to highlight the connection to amplitude amplification, we sketch the $N+1$ basis state amplitudes with colouring representing complex phase. In this perspective, $U_\text{OAT}$ applies a phase that depends non-linearly on the excitation number $M$, with the rotation then amplifying or attenuating the amplitude conditioned on the phase that was applied. Repeated application of $U_\text{OAT}$ and $R$ with optimized parameters can therefore maximize the amplitude of a target state. 
}
\label{fig:system}
\end{figure*}

\emph{Dicke states:} We consider an ensemble of $N$ qubits. The collective spin operators are defined as $J_\alpha = (1/2)\sum_{j=1}^{N} \sigma_j^\alpha$ where $\alpha \in \{x,y,z\}$ and $\sigma_j^\alpha$ is the Pauli operator acting on the $j$-th qubit. Dicke states, denoted $\ket{J,M}$, are simultaneous eigenstates of the total spin operator $J^2 = J_x^2 + J_y^2 + J_z^2$ and $J_z$. The quantum numbers $J$ and $M$ are defined by $J^2 \ket{J,M} = J(J+1)\ket{J,M}$, $J_z \ket{J,M} = M \ket{J,M}$ with $J \in \{0,\hdots,N/2\}$, and $M \in \{-J,\hdots,J\}$. We focus on the $N+1$ symmetric $J \equiv N/2$ Dicke states, which in the single-particle $\sigma_z$ basis are
\begin{align}
    \ket{J=N/2,M} = \frac{1}{\sqrt{\lambda}} \sum_{\mathcal{P}} ( \ket{1}^{J+M} \otimes \ket{0}^{J-M}),
\end{align}
where the normalisation factor is $\lambda = \begin{pmatrix}2J\\J+M\end{pmatrix}$, and where the summation is over all permutations $\mathcal{P}$ of the ordering of the spins \footnote{For example, the four $N=3$ symmetric Dicke states are $\ket{3/2,-3/2} = \ket{000}$, $\ket{3/2,-1/2} = 1/\sqrt{3}(\ket{001} + \ket{010} + \ket{100})$, $\ket{3/2,1/2} = 1/\sqrt{3}(\ket{011} + \ket{101} + \ket{110})$ and $\ket{3/2,3/2} = \ket{111}$}. Dicke states feature entanglement that is persistent~\cite{PhysRevLett.86.910} and robust to particle loss~\cite{wangPairwiseEntanglementSymmetric2002,stocktonCharacterizingEntanglementSymmetric2003,nevenEntanglementRobustnessParticle2018} which makes them attractive for experimental realization, while also demonstrating useful sensing capabilities~\cite{Wineland1994a,RevModPhys.90.035005}. Specifically, the $N$-qubit W-state $\ket{N/2,-N/2+1}$ is maximally pairwise entangled~\cite{koashiEntangledWebsTight2000,PhysRevA.62.062314}, while the Dicke state with $N/2$ excitations $\ket{N/2,0}$ saturates the quantum Cramér-Rao bound when sensing global rotations about the x-y axis~\cite{PhysRevLett.125.190403}. There are many existing deterministic preparation protocols for Dicke states, including quantum circuit approaches that use one- and two-qubit gates~\cite{9774323,9951196,buhrmanStatePreparationShallow2024}, global pulse schemes~\cite{PhysRevLett.125.190403,Ivanov_2013,stojanovicDickestatePreparationGlobal2023,PhysRevA.80.052302}, and others~\cite{PhysRevA.95.013845,higginsSuperabsorptionLightQuantum2014,zhaoCreationGreenbergerHorneZeilingerStates2021a}. 

In addition to bare Dicke states, superpositions of Dicke states are of key interest. For example, the Greenberger–Horne–Zeilinger (GHZ) state $\ket{\text{GHZ}} = \ket{N/2,-N/2} + \ket{N/2,N/2}$ is maximally entangled, represents a distinct class of multi-partite entanglement, exhibits phase sensing sensitivity below the standard quantum limit (SQL)~\cite{Greenberger1989,PhysRevA.54.R4649} and has been prepared in many experimental platforms~\cite{PhysRevLett.106.130506,doi:10.1126/science.aax9743,doi:10.1126/science.aay0600}. More generally, superpositions over more of the Dicke states form the logical basis states of various QEC codes including Pauli exchange error correcting codes~\cite{PhysRevLett.85.194} and large spin codes~\cite{PhysRevLett.127.010504}. The weighting of individual Dicke states in each of these logical basis states are chosen to satisfy the Knill-Laflamme QEC conditions \cite{knillTheoryQuantumErrorCorrecting2000}. Existing protocols to prepare such specific, highly entangled superpositions of Dicke states typically require single-site addressability \cite{10.1007/978-3-030-25027-0_9} or a large detuning limit~\cite{zouGenerationAnySuperposition2003}. Below, we alleviate these constraints and propose to prepare Dicke state superpositions using a global variational quantum circuit. 

\emph{State preparation protocol: } Global variational quantum circuits have been previously proposed~\cite{carrascoExtremeSpinSqueezing2022,kaubrueggerQuantumVariationalOptimization2021,kaubrueggerVariationalSpinSqueezingAlgorithms2019,gutman2023universal} and demonstrated experimentally~\cite{marciniakOptimalMetrologyProgrammable2022a} for the preparation of squeezed quantum states for quantum metrology. Here, instead of optimizing the variational parameters to maximize a property of the target state (such as the squeezing amplitude or phase sensitivity), we employ such circuits to prepare target symmetric quantum states by minimizing the infidelity $1-\mathcal{F} = 1 - |\braket{\psi_{\text{t}}}{\psi_{\text{f}}}|^2$, where $\ket{\psi_{\text{t}}}$ is a chosen target state, $\ket{\psi_{\text{f}}} = U_\text{tot}\ket{\psi_0}$ the final state and $\ket{\psi_0}$ the initial state. As sketched in Fig.~\ref{fig:system}, the circuit of $P$ layers is
\begin{align}
    U_\text{tot} = \prod_{k=1}^{P} [R(\theta_k,\xi_k) U_\text{OAT}(\phi_k)], \label{eq:GateSequence}
\end{align}
and consists of: (i) global rotations $R(\theta_k,\xi_k) = R_z(\theta_k) R_y(\xi_k)$, and (ii) global one-axis twisting $U_\text{OAT}(\phi) = e^{i \phi J_z^2 }$, where $\phi = g t$ with $g$ the coupling strength and $t$ the evolution time~\cite{Kitagawa1993}. Because $R_z$ commutes with $U_\text{OAT}$, for $P\geq2$ the global rotation $R$ is about an arbitrary axis. Although the fidelity can be efficiently computed classically because the symmetric space dimension is $N+1$, in the Supplemental Material we discuss how an upper bound on the infidelity can be efficiently computed on quantum devices for a hybrid classical-quantum algorithm approach, which may be a preferred option to mitigate certain hardware errors~\cite{supp}. 

The capability of the protocol to prepare superpositions of Dicke states can be understood using amplitude amplification, as sketched in Fig.~\ref{fig:system}. To emphasize the versatility of the variational circuit, we note that $U_\text{OAT}$ and $R$ provide universal control over the symmetric subspace~\cite{albertiniControllabilitySymmetricSpin2018,gutman2023universal}. Although a general explicit expression for the optimal parameters for arbitrary symmetric state synthesis is a complicated task, a simple consistency argument for the minimum number of gates $P^*$ to prepare a general symmetric state can be made as follows: accounting for the coefficients and normalization of the $N+1$ basis states, a general state in the symmetric $J=N/2$ subspace is described by $2N$ real parameters. As each protocol step is characterized by three free parameters $\theta, \xi, \phi$, the preparation of an arbitrary symmetric state requires $P^* = 2N/3$ steps. 

We numerically verify the scaling relation for $P^*$ by generating $N_\text{Haar} = 200$ Haar random states as target states and optimizing the protocol's variational parameters to minimize $1-\mathcal{F}$. Obtaining (and verifying) a global minima is challenging, as in general the optimization landscape can contain many local minima. However, if the system is controllable~\cite{NoteControllable}, then every local minima is a global minima~\cite{rabitzQuantumOptimallyControlled2004}. As such, performing only local optimization of the parameters we expect a sharp transition in target state infidelity at some $P \geq P^*$ as the system becomes controllable.

\begin{figure}[t]
    \centering
    \includegraphics[width=0.9\linewidth]{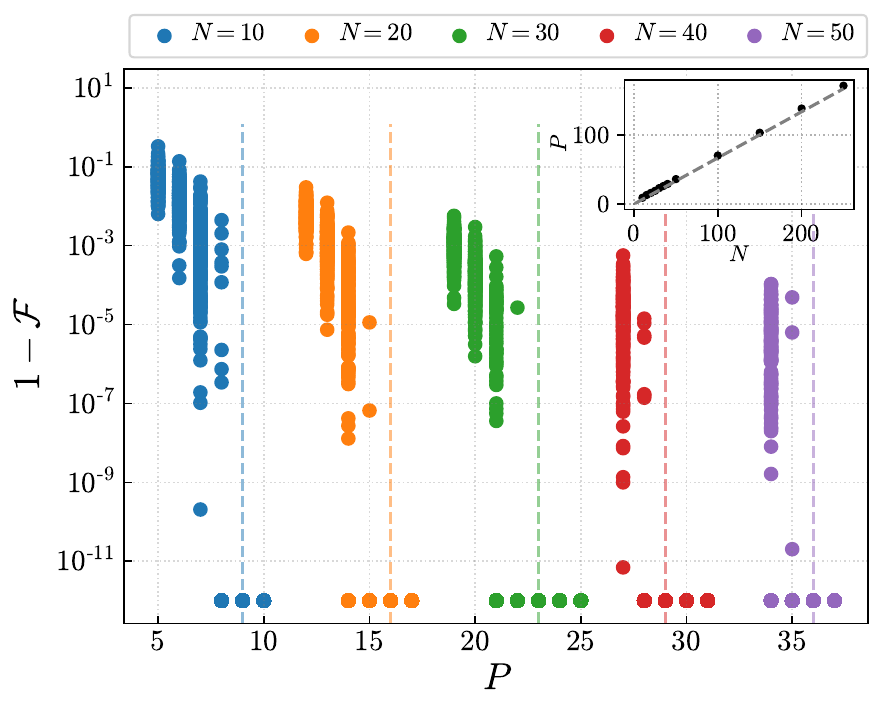}%
    \caption{
    Infidelity $1-\mathcal{F} = 1 - |\braket{\psi_{\rm t}}{\psi_{\rm f}}|^2$ vs. circuit depth $P$ for various $N$. Each data point is a Haar random target state; we draw $N_\text{Haar} = 200$ targets for each $N$. The variational parameters are classically optimized using a local optimization algorithm. As the system becomes controllable, all local minima become global minima, which corresponds for each $N$ to the sharp decrease in infidelity as $P$ increases (marked with vertical dashed lines). Inset: location of the transition as a function of $N$ (black points) for up to $N = 250$, demonstrating excellent agreement with the scaling prediction $P^* = 2N/3$ (grey dashed). 
    }
    \label{fig:InfidelityVsP_HaarRandom}
\end{figure}

In Fig.~\ref{fig:InfidelityVsP_HaarRandom} we plot the infidelity $1-\mathcal{F}$ as a function of $P$ for various $N$ up to $N = 50$. We perform local gradient-based optimisation using \textsc{BFGS}~\cite{nocedalNumericalOptimization2006} in \textsc{Julia}~\cite{Julia-2017} with \textsc{Optim.jl} and \textsc{LineSearches.jl}~\cite{mogensen2018optim}, and terminate the optimization once the infidelity passes below a threshold $\epsilon = 10^{-12}$. As expected, as $P$ increases the infidelity decreases, with a relatively sharp transition as the system becomes controllable. In the inset of Fig.~\ref{fig:InfidelityVsP_HaarRandom} we plot the location of the transition as a function of $N$, showing excellent agreement with the scaling obtained from counting, $P^* = 2N/3$. As shown in the inset we further verify this prediction by optimizing the parameters for $N_{\rm Haar} = 200$ Haar random states using $P =  \ceil{2N/3} + c$ gates with a small constant offset $c$~\footnote{An offset $c$ in the number of gates $P = \ceil{2N/3}+c$ is needed to reach fidelities below the numerical termination threshold $\epsilon = 10^{-12}$. For $N = 10$ to $N = 150$ the offset is either $c = 2$ or $c = 3$, depending on the ceil function. For $N = 200$ and $N = 250$ the offset is $c = 4$, to give the numerical optimization more overhead to reach the termination threshold. Specifically, for $N = [100,150,200,250$ we use $P = [70,103,138,171]$ gates.} for $N = 100$, $N = 150$, $N = 200$ and $N = 250$, verifying that all states can be prepared with infidelities below the optimization termination threshold $\epsilon = 10^{-12}$. Therefore the protocol Eq.~(\ref{eq:GateSequence}) is an efficient preparation protocol for synthesizing arbitrary symmetric states. To the best of our knowledge this improves upon previous global control protocols, which required $N$ steps only in a large detuning limit~\cite{zouGenerationAnySuperposition2003} or using excited Rydberg states \cite{PhysRevLett.117.213601}, and $\mathcal{O}(N^4)$ steps for arbitrary unitary synthesis~\cite{gutman2023universal}. 

\emph{Dicke state preparation: } Next, we demonstrate the preparation of specific states, beginning with bare Dicke states. Because the gate sequence Eq.~(\ref{eq:GateSequence}) starts with the application of $U_\text{OAT}$, for the initial state we choose a global rotation acting on $\ket{N/2,-N/2}$, which is a coherent spin state (CSS) $\ket{\psi_0} = \ket{\bar{\theta},\bar{\phi}}$, i.e. an eigenstate of the spin component in the $(\bar{\theta},\bar{\phi})$ direction. The parameters $(\bar{\theta},\bar{\phi})$ are also numerically optimized, and thus the total number of free parameters is $3P+2$. Both here and in the next section, the numerical optimisation uses a multi-start Monte Carlo search, with local optimization using \textsc{L-BFGS}~\cite{liuLimitedMemoryBFGS1989} 

\begin{figure}[t]
    \centering
    \includegraphics[width=0.9\linewidth]{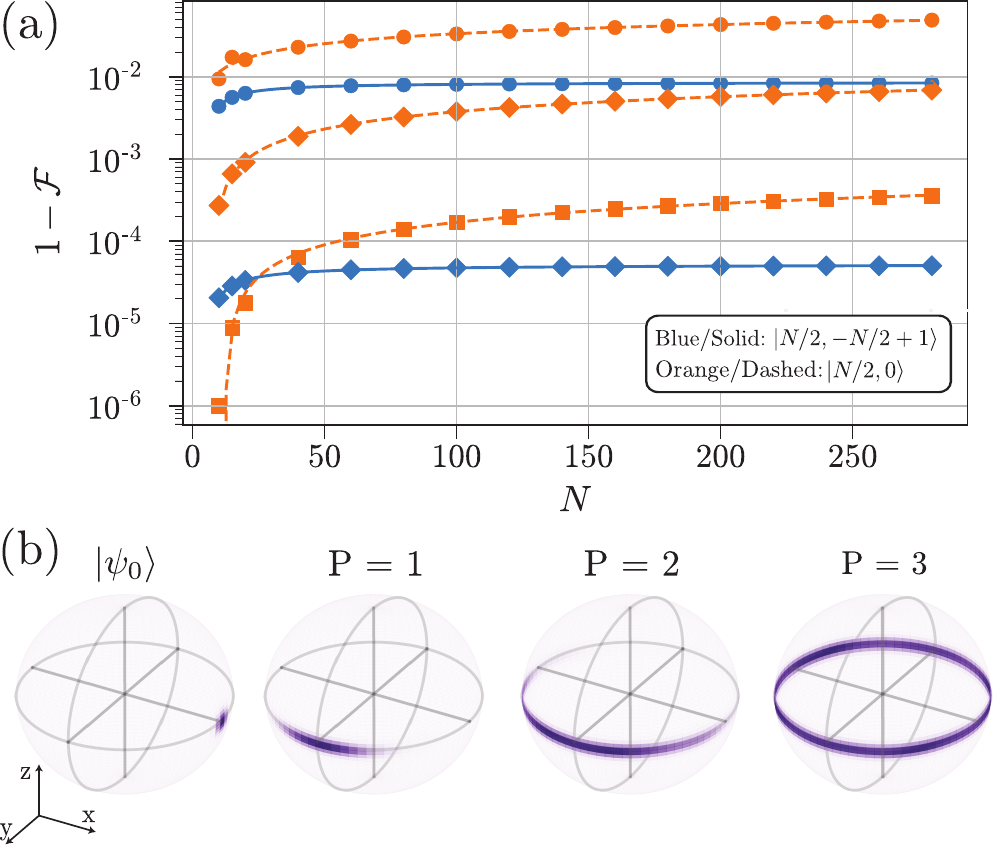}%
    \caption{(a) Theoretical infidelity $1-\mathcal{F} = 1-|\braket{\psi_{\text{t}}}{\psi_{\text{f}}}|^2$ to prepare: the the W state $\ket{N/2,-N/2+1}$ (solid blue) and $N/2$ excitation state $\ket{N/2,0}$ (dashed orange) using the protocol $U_\text{tot}$ Eq.~(\ref{eq:GateSequence}) for $P = 2$ (dots), $P = 3$ (diamonds) and $P = 4$ (squares). The lines show a power law fit, with excellent agreement. Both target states can be prepared with small theoretical infidelities $1-\mathcal{F} < 10^{-3}$ in $P = 3$ and $P = 4$ gate steps respectively in systems as large as $N = 300$ qubits. (b) The su(2) Husimi Q-representation for the $P = 3$ solution that prepares $\ket{N/2,0}$ with $N = 300$. The optimized initial coherent spin state starts along $x$, with the subsequent interleaving of one-axis twisting and global rotations leading to the preparation of $\ket{\psi_t}$ with $1-\mathcal{F} \sim 10^{-2}$. 
    }
    \label{fig:OptimResults}
\end{figure}

Firstly, we prepare the Dicke state with $N/2$ excitations, $\ket{\psi_{\text{t}}} = \ket{N/2,0}$. This state is metrologically useful: when the state used for sensing is $\ket{N/2,0}$, the uncertainty in the measured angle $\epsilon$ of a rotation $e^{-i \epsilon J_y}$ saturates the Cramer-Rao bound, $(\Delta \epsilon)^2 = 2/[N(N+2)]$ and features entanglement that is robust to particle loss~\cite{Apellaniz_2015,9635793,wangPairwiseEntanglementSymmetric2002,stocktonCharacterizingEntanglementSymmetric2003,nevenEntanglementRobustnessParticle2018}. 
In Fig.~\ref{fig:OptimResults}(a) we plot (dashed orange) $1-\mathcal{F}$ as a function of $N$ for $P = 2$ (circles), $P = 3$ (diamonds) and $P = 4$ (squares). The infidelity is remarkably robust to increasing qubit number, with $P = 4$ gates sufficient to prepare the target state in up to $N = 300$ qubits with $1 - \mathcal{F} \lesssim 10^{-2}$. The dashed lines show a power law fit with excellent agreement, emphasizing the slow increase of infidelity with $N$ for $N \gtrsim 50$. In Fig.~\ref{fig:OptimResults}(b) we plot the evolution of the $N = 300$ state under the $P = 3$ gate sequence solution using the su(2) Husimi Q-Representation $Q(\theta,\phi) = \bra{\theta,\phi} \rho \ket{\theta,\phi}$ of a state $\rho$ where $\ket{\theta,\phi}$ is a CSS. Clearly visible is the effect of spin squeezing, whose repeated application combined with global rotations results in the preparation of $\ket{N/2,0}$.

Secondly, we prepare the $N$-qubit W state, $\ket{N/2,-N/2+1}$. In Fig.~\ref{fig:OptimResults}(a), we plot (solid blue) the infidelity to prepare the  W-state with $P = 2$ (circles) and $P = 3$ (diamonds)~\footnote{The superposition of all states with only one qubit not excited $\ket{N/2,N/2-1}$ can be prepared using the same sequence by applying a global spin flip at the end of the sequence, and that optimizing for $\ket{N/2,N/2-1}$ rather than $\ket{N/2,-N/2+1}$ does not offer any potential for improvement as the initial state is an optimized CSS.}. We find that the W state can be prepared in systems of up to $N = 300$ qubits with small infidelities $1-\mathcal{F} < 10^{-4}$ in only $P = 3$. Thus, our protocol can be used to prepare these two Dicke states in up to $N = 300$ qubits in $\mathcal{O}(1)$ gate steps. 

\emph{Quantum error correction codewords: }
Next, we prepare specific superpositions of Dicke states for QEC. Pauli exchange errors are two-qubit errors that arise due to interactions between qubits, for example due to dipole-dipole interactions. Physically, exchange errors are similar to bit flip errors, with the additional constraint that a qubit will flip if and only if its neighbour is different. Codes such as the $9$-qubit Shor code~\cite{shor1995scheme} cannot distinguish between exchange errors and phase errors. The Ruskai code~\cite{PhysRevLett.85.194} exploits the permutation invariance of Dicke states to correct all one qubit errors and Pauli exchange errors. The explicit codewords for $N = 9$ are
\begin{subequations}
\begin{align}
    \ket{R_0} &= \ket{9/2,-9/2}/2 + \sqrt{3/4}\ket{9/2,3/2}, \\ 
    \ket{R_1} &= \ket{9/2,9/2}/2 + \sqrt{3/4}\ket{9/2,-3/2},
\end{align}
\end{subequations}

As a second example, we consider the Gross codes \cite{PhysRevLett.127.010504}. Designed such that the logical single-qubit Clifford operations are global rotations, the Gross code is immune, to first order, to errors that correspond to global rotations. As such, the code utilizes interactions that are native to many experimental platforms, and protects against the typically most deleterious sources of noise, including both $T_1$-type relaxation and $T_2$-type dephasing errors. The smallest collective spin for which this code can be constructed is $N = 13$. It utilizes the two following states $\ket{\tilde{G}_{0,1}} = c_1 \ket{13/2,13/2} + c_2 \ket{13/2,5/2} + c_3 \ket{13/2,-3/2} + c_4 \ket{13/2,-11/2}$, where the coefficients for $\ket{\tilde{G}_0}$ are $c_1 = \sqrt{910}/56$, $c_2 = -3\sqrt{154}/56$, $c_3 = -\sqrt{770}/56$ and $c_4 = \sqrt{70}/56$; and for $\ket{\tilde{G}_1}$ are $c_1 = \sqrt{231}/84$, $c_2 = \sqrt{1365}/84$, $c_3 = -\sqrt{273}/28$ and $c_4 = -\sqrt{3003}/84$. The explicit codewords are then
\begin{align}
    \ket{G_{0}} = \frac{\sqrt{105}}{14}\ket{\tilde{G}_0} + \frac{\sqrt{91}}{14}\ket{\tilde{G}_1},
\end{align}
where $\ket{G_1}$ follows from $\ket{G_0}$ by a global spin flip. 

In Fig.~(\ref{fig:QECStatePreparation}) we plot the infidelity of the state $\ket{\psi_{\text{f}}}$ prepared using the optimized gate protocol Eq.~(\ref{eq:GateSequence}) with target states $\ket{\psi_{\text{t}}} = \ket{R_1}$ (blue with dots) and $\ket{\psi_{\text{t}}} = \ket{G_0}$ (orange with stars). We find solutions that prepare the codewords with infidelities  $1-\mathcal{F} < 10^{-4}$ in only $P = 4$ and $P = 7$ gate steps, respectively. Therefore, our protocol provides a pathway towards the implementation of large $N$-qubit QEC codes utilizing only global addressability of the qubits constituting the codeword. 

\begin{figure}
    \centering
    \includegraphics[width=0.9\linewidth]{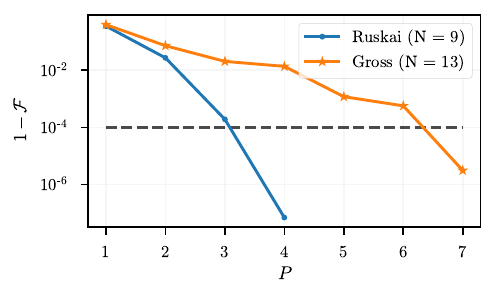}
    \caption{Infidelity to prepare the $N = 9$ Ruskai $\ket{R_1}$ and $N = 13$ Gross $\ket{G_0}$ codewords using the optimized gate sequence as a function of $P$. Despite being highly specific superpositions of Dicke states, we numerically find solutions that can prepare both target states with infidelities $1 - \mathcal{F} < 10^{-4}$ in $P = 4$ and $P = 7$ gate steps for $\ket{R_1}$ and $\ket{G_0}$, respectively.}
    \label{fig:QECStatePreparation}
\end{figure}

\emph{Realization in trapped ions:}
$U_\text{OAT}$ can be realized in trapped ions by off-resonantly driving the centre of mass (COM) mode~\cite{Molmer:1999,Kim:2009,Britton:2012,PhysRevLett.106.130506,Zhang:2017}. Considering $N$ ions in a linear Paul trap, two global Raman beams detuned from the red and blue sidebands by $\delta$ leads to the following Hamiltonian in the interaction picture,
\begin{align}
    H_I = \Omega \eta (a e^{-i \delta t} + a^\dagger e^{i \delta  t} ) J_z, \label{eq:HI}
\end{align}
where we have assumed the Lamb-Dicke regime, and where $a$ ($a^\dagger$) are the annihilation (creation) operators of the COM mode, $\eta$ the Lamb-Dicke parameter and $\Omega$ the Rabi frequency. Note that although we require the Lamb-Dicke regime, our protocol is otherwise insensitive to temperature and therefore does not require ground state cooling~\cite{Molmer:1999,Kirchmair_2009}. Evolving for time $t_f = 2\pi/\delta$ results in a closed loop in phase space such that the motion decouples from the spin degree of freedom. The effective evolution is described by one-axis twisting, $U_\text{OAT} = \exp(i \phi J_z^2)$ with $\phi = 2\pi \eta^2 \Omega^2/\delta^2$ the area enclosed~\cite{Sorensen2000}. Our optimised solutions are typically $\phi \sim 1$. Setting $\eta \Omega = 2\pi \times 20$~kHz and $\delta = 2\pi \times 20$~kHz results in $t_\text{f}=50$~$\mu$s and $\phi = 2\pi$. The total protocol duration $T$ is therefore $T = P t_\text{f}$, and thus for all the gate sequences we present $T \lesssim 350$~$\mu$s. 

Here we study the effects of off-resonant driving and dephasing, with laser intensity fluctuations and non-global beams further analysed in the Supplemental Material~\cite{supp}. Off-resonant driving of the other modes leads to non-uniform spin-dependent phases. Assuming a typical trap frequency $\omega = 2\pi \times 2$~MHz, the stretch mode, which is the leading order correction, is detuned by $(\sqrt{3}-1) \omega + \delta$. The area enclosed by the stretch mode is $\sim10^{-2}$ times smaller than the area enclosed by the COM mode, and thus for realistic parameters unwanted off-resonant driving of the collective modes is negligible compared to the effects of dephasing, which we consider below.

We use a realistic dephasing rate $\gamma = 2 \pi \times 5$~Hz. The off-diagonal coherences of the GHZ state decay as $\sim e^{- \gamma N t}$, such that the fidelity can be roughly estimated as $F \sim 1/2 ( 1 + e^{-\gamma N t})$. We assume that the rotations are noise-free, as they can be performed fast relative to the spin squeezing. For the $N = 9$ Ruskai and $N= 13$ Gross protocols, using the $P = 4$ and $P = 7$ gate sequences respectively, assuming the ideal protocol prepares the state with perfect fidelity, we estimate noisy fidelities $\mathcal{F} \sim 97\%$ and $\mathcal{F} \sim 93\%$ respectively. Reducing the dephasing rate to $\gamma = 2\pi \times 1$Hz increases the Ruskai and Gross codeword fidelities to $\mathcal{F} \sim 99\%$ and $\mathcal{F} \sim 98\%$ respectively. Furthermore, parametric amplification can increase the squeezing amplitude achieved in a given $t_\text{f}$~\cite{PhysRevLett.122.030501}, whilst multiple Raman beatnotes can reduce off-resonant driving enabling a larger $\delta$~\cite{Shapira:2020,Espinoza:2021b}. Such techniques, when combined with achievable reductions in dephasing noise, could bring state preparation infidelities of the Gross and Ruskai codewords below QEC thresholds in the near-term. 

\emph{Outlook: }
We showed that global variational circuits of linear depth can prepare arbitrary superpositions of Dicke states. The circuit can be easily realized across a variety of experimental platforms. Its implementation could lead to the preparation of interesting and useful entangled states, with applications in quantum metrology and quantum error correction. 

Finally, we note that the set of realizable interactions depends on the experimental platform, e.g. in cavity QED experiments dispersive interactions can be used to engineer the generator $ \sin(J_z + \varphi) $~\cite{PhysRevLett.125.190403}, whilst addressing the second red and blue sidebands in trapped ions can produce $ \cosh(r J_z) + \cos(\varphi) \sinh(r J_z)$. These Hamiltonians can replace $U_\text{OAT}$; or the inclusion of more than one flavour of phase gate in the protocol could provide additional geometric control, potentially reducing the total number of gate steps required. 

\emph{Acknowledgments: } 
We thank Dominic Berry for discussions. This work was supported by the Netherlands Organization for Scientific Research (Grant Nos. OCENW.XL21.XL21.122 and VI.C.202.051, (R.G.)), and the Australian Research Council (ARC) Centre of Excellence for Engineered Quantum Systems (EQUS, CE170100009). A.S.N is supported by the Dutch Research Council (NWO/OCW), as part of the Quantum Software Consortium programme (project number 024.003.037) and Quantum Delta NL (project number NGF.1582.22.030).

\bibliography{library}

\begin{thebibliography}{73}
\bibitem [{Not()}]{NoteControllable}%
  \BibitemOpen
  \href@noop {} {}\bibinfo {note} {If there exists a control field $C(t)$ such that $\delta \mathcal{P}_{i \rightarrow f}/\delta C(t) = 0$ $\forall\, 0 \leq t \leq T$, where $\mathcal{P}_{i \rightarrow f} = |\bra{\psi_f} U_{\rm tot} \ket{\psi_0}|^2$ is the transition element for a transition from an initial (final) state $\ket{\psi_0}$ ($\ket{\psi_f}$) for some finite protocol time $T$, then the system is controllable.}\BibitemShut {Stop}%
\end{thebibliography}


\ifincludeSupplement

\clearpage

\setcounter{equation}{0}
\setcounter{figure}{0}
\setcounter{table}{0}
\setcounter{page}{1}

\renewcommand\thefigure{\arabic{figure}}

\let\theequationWithoutS\theequation 
\renewcommand\theequation{S\theequationWithoutS}
\let\thefigureWithoutS\thefigure 
\renewcommand\thefigure{S\thefigureWithoutS}

\title{Supplemental Material: \mytitle}

\maketitle

\onecolumngrid

\section{Errors due to laser intensity fluctuations}
In this section we study the effect of laser intensity fluctuations on the protocol given in Eq.~(\ref{eq:GateSequence}) of the main text. We focus on an implementation in trapped ions, and consider worst-case shot-to-shot noise. We assume that the noise is proportional to the one-axis twisting duration, and simulate the noise by multiplying each one-axis twisting parameter $\phi_k$ by a fluctuation randomly sampled from a Gaussian distribution with standard deviation $\delta \phi \times \phi_k$, where $\delta \phi$ is a scaling factor~\cite{bondEffectMicromotionLocal2022a}. We obtain an average infidelity by averaging the noisy protocol infidelity over 200 realizations of the random Gaussian sampling. In Fig.~\ref{fig:Noisy} we plot the resulting averaged infidelity as a function of $\delta\phi$ for (a) one Haar random state for various values of $N$, and (b) the $N=9$ Ruskai and $N=13$ Gross codewords. For the sake of demonstration, the Haar random state is chosen from the $N_{\rm Haar} = 200$ random states that were generated for Fig.~\ref{fig:InfidelityVsP_HaarRandom} of the main text by choosing the state whose protocol parameters had the smallest total one-axis twisting amplitude, i.e. the smallest $\Phi = \sum_i |\phi_i|$. As expected, as both $N$ and $\delta \phi$ increase the infidelity increases until saturation $1-{\cal F}=1$ at large $\delta \phi$. We note that the infidelity of both codewords remains low, $1-\mathcal{F} \lesssim 2 \times 10^{-4}$ and $1-\mathcal{F} \lesssim 2 \times 10^{-3}$ for Ruskai and Gross respectively, even up to fluctuations as large as $ \delta \phi = 0.1\%$. 

\begin{figure}
\centering
\subfloat{%
  \includegraphics[width=0.49\columnwidth]{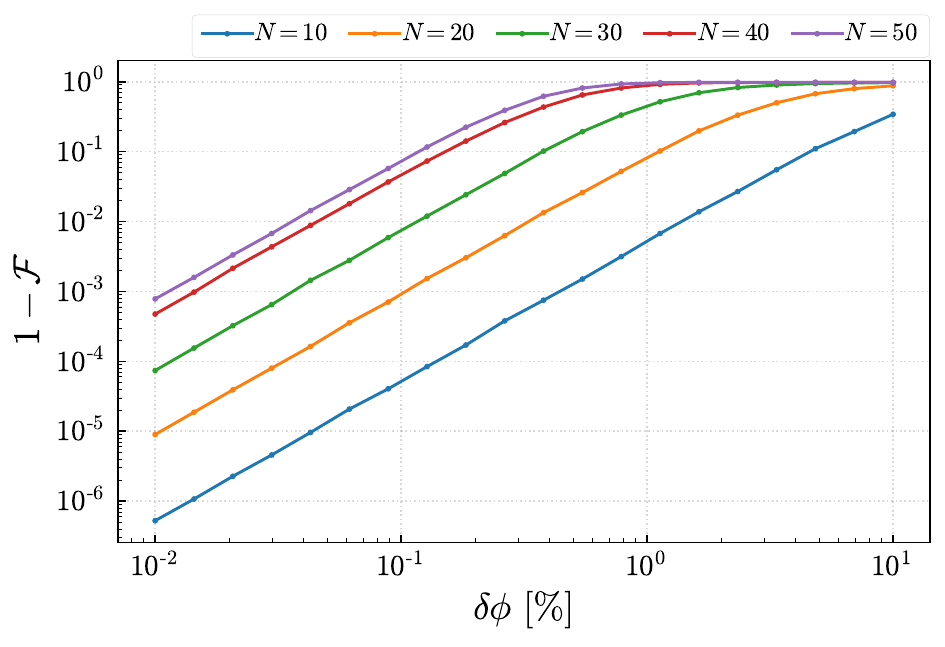}%
}\hfill
\subfloat{%
  \includegraphics[width=0.49\columnwidth]{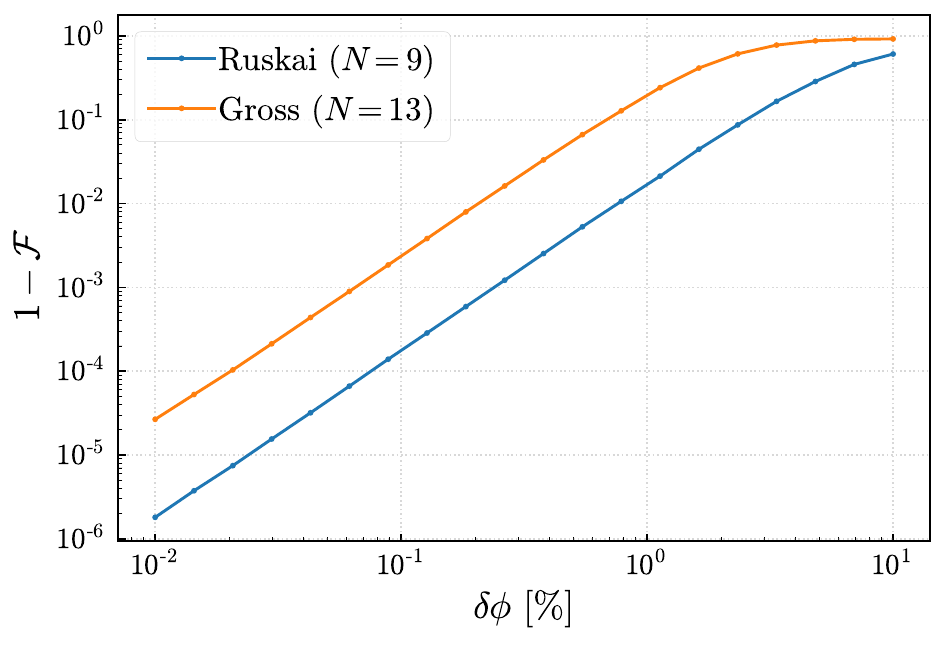}%
}\hfill
\caption{Averaged infidelity including shot-to-shot laser intensity fluctuations during the preparation of (a) one Haar random state and (b) Ruskai and Gross codewords. The scaling factor $\delta \phi$ enters the standard deviation $\delta \phi \times \phi_k$ of the Gaussian from which the noisy one-axis twisting parameter $\phi_k$ is sampled. The infidelity is proportional to $\delta \phi$ and $N$.}
\label{fig:Noisy}

\end{figure}

\section{Errors due to non-global M\o{}lmer-S\o{}rensen gate}
In this section we consider the effect of non-global one-axis twisting. As discussed in Eq.~(\ref{eq:HI}) of the main text, in trapped ions one-axis twisting is realized with a M\o{}lmer-S\o{}rensen interaction. If the laser intensity of the Raman beam is nonuniform, the Rabi frequency $\Omega$ becomes site-dependent, so that the interaction Hamiltonian Eq.~(\ref{eq:HI}) becomes 
\begin{align}
        H_I'(\Omega_i) = \eta (a e^{-i \delta t} + a^\dagger e^{i \delta  t} ) \sum_i \frac{\Omega_i \sigma_i^z}{2}. 
\end{align}
We simulate the effect of a non-uniform Raman beam by drawing $\Omega_i$ from a Gaussian function of standard deviation $\sigma$ at $x$-values corresponding to the dimensionless positions of the ions $u_i = x_i/\ell$, as shown in Fig.~\ref{fig:nonglobalerror}(a)-(b), where the length scale is $\ell^3 = Z^2 e^2/(4\pi\epsilon_0 M \omega^2)$, with $Z$ the degree of ionization, $e$ the electron charge, $\epsilon_0$ the free space permittivity, $M$ the ion mass and $\omega$ the trap frequency. We fix $\eta = 1$, $\Omega = 2\pi \times 20~\text{kHZ}$ and vary the detuning $\delta$ so as to realize the desired one-axis twisting angle $\phi$ according to $\phi = 2\pi \eta^2 \Omega^2/\delta^2$. In panel (c) we plot the resulting infidelity as a function of $\sigma$ for the preparation of the middle Dicke state $\ket{J=N/2,0}$ for $N = 6$, $N = 8$ and $N = 10$. As expected the infidelity increases with increasing $N$, as seen in the order of magnitude increase in infidelity when going from $N = 6$ to $N = 10$ for a given standard deviation value $\sigma$. The infidelity also increases with the total one-axis twisting duration $\Phi = \sum_i \phi_i$, as seen in the similarities between the $N = 8$ and $N = 10$ curve, which have $\Phi = 17.22$ and $\Phi = 12.71$, respectively. Thus, finding optimal parameters that minimize the total one-axis twisting duration is crucial for minimizing the effect of non-global one-axis twisting.

\begin{figure}
    \centering
    \includegraphics[width=0.8\linewidth]{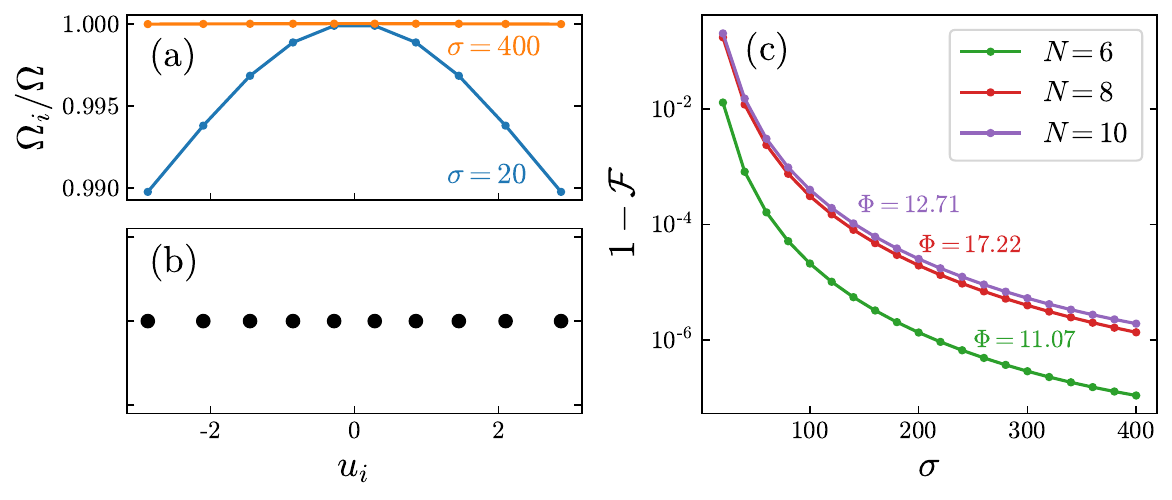}
    \caption{(a) Gaussian distribution of standard deviation $\sigma$ for $\Omega_i$, with the dots showing the value for $\Omega_i$ sampled at the dimensionless positions of the ions $u_i$ for $N = 10$. (b) Equilibrium positions for a $N=10$ ion chain. (c) Infidelity vs Gaussian standard deviation $\sigma$. The infidelity increases with increasing $N$ and increasing total one-axis twisting amplitude $\Phi$. }
    \label{fig:nonglobalerror}
\end{figure}

\section{Hybrid quantum-classical algorithm}
In this section we provide details on how the variational circuit of Eq.~\eqref{eq:GateSequence} of the main text can be implemented as a hybrid quantum-classical algorithm~\cite{kaubrueggerQuantumVariationalOptimization2021,marciniakOptimalMetrologyProgrammable2022a,mccleanTheoryVariationalHybrid2016,peruzzoVariationalEigenvalueSolver2014,kaubrueggerQuantumVariationalOptimization2021,kaubrueggerVariationalSpinSqueezingAlgorithms2019}. Specifically we consider how the cost function (infidelity) can be efficiently measured on a quantum device and then fed to the classical optimizer. The advantage of this approach is that the optimization is noise-aware, potentially enabling the mitigation of certain types of noise. For example, a naive strategy to minimize the effect of dephasing on the fidelity is to minimize the total one-axis twisting duration. The hybrid quantum-classical approach natively incorporates the effect of dephasing into the cost function by evaluating the infidelity on-device, such that the classical optimization algorithm can balance finding the (noiseless) ideal parameters with minimizing the total one-axis twisting gate time. 

For the following, we assume that although the experimenter only has global control over the qubits, they have individual qubit readout, as is often the case in platforms such as trapped ions. Then, permutationally invariant (PI) tomography can be used to compute the fidelity of the prepared state with respect to the target state using $\mathcal{O}(N^2)$ global measurement settings, where a global measurement setting is defined as a choice of single-qubit observable $A_j = a_{j,x} \sigma_x + a_{j,y} \sigma_y + a_{j,z} \sigma_z$ measured simultaneously (globally) on all $N$ qubits, i.e. $A_j^{\otimes N}$. Below we briefly review PI quantum tomography following Ref.~\cite{tothPermutationallyInvariantQuantum2010}. 

Firstly, the PI density matrix is defined as 
\begin{align}
    \rho_{\rm PI} = \frac{1}{N!}\sum_k \Pi_k \rho \Pi_k, \label{eq:rhoPI}
\end{align}
where $\Pi_k$ is the qubit permutation operator and $\rho$ the density matrix. A symmetric state $\rho$ is invariant under the action of $\Pi_k$ and thus $\rho=\rho_{\rm PI}$. To obtain $\rho_{\rm PI}$ from experimentally accessible measurements, note that $\rho_{\rm PI}$ can be written as a linear combination of $\mathcal{D}_N = \begin{pmatrix}N+2\\N\end{pmatrix} = \frac{1}{2}(N^2+3N+2)$ operators of the form $(\sigma_x^{\otimes k} \sigma_y^{\otimes l} \sigma_z^{\otimes m} \mathbb{1}^{\otimes n})_{\rm PI}$, where $k+l+m+n=N$ and where the $(\cdot)_{\rm PI}$ notation indicates an operator of the form of Eq.~(\ref{eq:rhoPI}). To avoid measuring $(\sigma_x^{\otimes k} \sigma_y^{\otimes l} \sigma_z^{\otimes m} \mathbb{1}^{\otimes n})_{\rm PI}$ directly which would require single-qubit control, note the decomposition~\cite{tothPermutationallyInvariantQuantum2010}
\begin{align}
    \langle (\sigma_x^{\otimes k} \sigma_y^{\otimes l} \sigma_z^{\otimes m} \mathbb{1}^{\otimes n})_{\rm PI} \rangle = \sum_j^{\mathcal{D}_N} c_j^{(k,l,m)} \langle (A_j^{\otimes(N-n)} \otimes \mathbb{1}^{\otimes n})_{\rm PI} \rangle, \label{eq:XYZDecomposition}
\end{align}
where $c_j^{(k,l,m)}$ are real coefficients and $A_j^{\otimes (N-n)}$ the $j$-th measurement setting acting on $(N-n)$ qubits. The choice of coefficients $c_j^{(k,l,m)}$ and measurement settings $A_j$ is not arbitrary. The optimal choices can be found using numerical optimization techniques as detailed in Ref.~\cite{tothPermutationallyInvariantQuantum2010}. 

Next, observe that the choice of measurement setting $A_j$ appearing in the RHS of Eq.~\ref{eq:XYZDecomposition} is independent of $k$, $l$ and $m$. For a given $j$, all of the expectation values $\langle (A_j^{\otimes(N-n)} \otimes \mathbb{1}^{\otimes n})_{\rm PI} \rangle$ for $0 \leq n \leq N$ can be reconstructed in post-processing from the coincidence counts of the global measurement setting $A_j^{\otimes N}$. The quantities to be measured appearing on the right-hand side are therefore entirely independent of the coefficients $k$, $l$, $m$ and $n$ appearing on the left-hand side. 

As such, for all values of $k,l,m$ and $n$, $\langle (\sigma_x^{\otimes k} \sigma_y^{\otimes l} \sigma_z^{\otimes m} \mathbb{1}^{\otimes n})_{\rm PI} \rangle$  can be reconstructed from the coincidence counts of an optimized set of $\mathcal{D}_N$ global measurement settings $\{A_j^{\otimes N}\}_{j=1}^{\mathcal{D}_N}$ with optimized coefficients $c_j^{(k,l,m)}$. Therefore, $\rho_{\rm PI}$ can be obtained using $\mathcal{D}_N$ global measurement settings. 

Finally, we note that in the case of leakage from the PI density matrix, i.e. due to any non-global operations or decoherence sources, the fidelity between the target PI state $\rho_{\rm PI}$ and the experimentally prepared state $\rho$ can be efficiently lower bounded~\cite{tothPermutationallyInvariantQuantum2010}. 

\subsection{Fidelity proxy}
Although the number of measurement settings required to perform PI tomography scales efficiently (quadratically) with $N$, here we discuss how under certain assumptions, a proxy for the fidelity cost function can be obtained using only a single measurement setting. Firstly, we assume that the target state $\ket{\psi_t}$ can be written in the $J=N/2$ Dicke state basis with real coefficients,
\begin{align}
    \ket{\psi_t} = \sum_{i=-J}^J b_i \ket{J,i}, \qquad b_i \in \mathbb{R}. \label{eq:DSBasis_Real}
\end{align}
Many states of interest are of the form of Eq.~(\ref{eq:DSBasis_Real}), including any single Dicke state and both Gross and Ruskai codewords~\cite{PhysRevLett.127.010504,PhysRevLett.85.194}. For a given Dicke state $\ket{J,m}$ we denote its corresponding density matrix $\rho_{J,m} = \ketbra{J,m}{J,m}$, which has a one at position $(m+N/2+1,m+N/2+1)$ and zeros elsewhere.

Next, we define a proxy for the fidelity, 
\begin{align}
        \mathbb{F}(\rho_{\rm exp},\rho_t) = \sum_{i=-J}^{J} \Tr{\rho_{\rm exp} \rho_{J,i}}\Tr{\rho_t \rho_{J,i}}, \label{eq:FidelityProxy}
\end{align}
where $\rho_{\rm exp}$ is the experimentally prepared state and $\rho_t = \ketbra{\psi_t}{\psi_t}$ the target state. In words, $\mathbb{F}$ measures the inner product between the vectors $(\Tr{\rho_{\rm exp} \rho_{J,i=-J}},\ldots,\Tr{\rho_{\rm exp} \rho_{J,i=J}})$ and $(\Tr{\rho_{t} \rho_{J,i=-J}},\ldots,\Tr{\rho_{t} \rho_{J,i=J}})$ which correspond to the decomposition of $\rho_{\rm exp}$ and $\rho_t$ on the Dicke state basis. Although $\mathbb{F}$ does not provide any information about the overall coherence of the quantum state, it does give information about our ability to coherently prepare a specific Dicke state $\ket{J,i}$ (which in the computational basis can be highly entangled) with target population $|b_i|^2$. Using the fact that $\rho_{J,m}$ can be written as a sum over powers of $J_z$ as $\rho_{J,m} = \sum_{i=0}^N c_{i,m} J_z^i$ for coefficients $c_{i,m} \in \mathbb{R}$ (for proof see below), the fidelities of $\rho_{\rm exp}$ with respect to each $\rho_{J,m}$ can be written as
\begin{align}
    \Tr{\rho_{\rm exp} \rho_{J,m}} = \sum_{i=0}^N c_{i,m} \Tr{J_z^i \rho_{\rm exp}}.
\end{align}
The expectation values $\Tr{J_z^i \rho_{\rm exp}}$ can in turn be decomposed using~\cite{tothPracticalMethodsWitnessing2009}
\begin{align}
    J_\alpha^n = \sum_{i=0}^n d_{i,\alpha} \sum_{k} \left[\Pi_k \sigma_\alpha^{\otimes i} \otimes \mathbb{1}^{\otimes (N-i)} \Pi_k\right], \qquad \alpha \in \{x,y,z\}, \label{eq:JlnDecomposition}
\end{align}
where the summation over $k$ is over all permutations of the qubit ordering, and where the coefficients $d_{i,\alpha}$ are defined by the expansion. For example, for $\alpha = x, n = 3$ we have $d_{0,1,2,3,\alpha} = 0,7/8,0,6/8$ because $J_x^3 = \frac{1}{8} (\sigma_x^{(1)} + \sigma_x^{(2)} + \sigma_x^{(3)})^3 = \frac{6}{8} \sigma_x^{(1)} \sigma_x^{(2)} \sigma_x^{(3)} + \frac{7}{8} (\sigma_x^{(1)} + \sigma_x^{(2)} + \sigma_x^{(3)})$ where $\sigma_x^{(i)}$ acts on the $i$th qubit. 

The RHS of Eq.~(\ref{eq:JlnDecomposition}) can always be reconstructed from the coincidence counts of a single measurement setting $\sigma_z^{\otimes N}$. Thus the populations $\Tr{\rho_{\rm exp} \rho_{J,m}}$ for all $m \in \{-J,\dots,J\}$, and therefore fidelity proxy $\mathbb{F}$ of Eq.~(\ref{eq:FidelityProxy}) can be obtained from a single measurement setting $\sigma_z^{\otimes N}$. 

A limitation of $\mathbb{F}$ is that, although it provides information about our ability to coherently prepare the state $\ket{J,i}$ with weight $|b_i|^2$, it does not give any information about the coherence of the overall superposition. For example, perhaps instead of preparing the coherent superposition $\ket{\psi} = 1/\sqrt{|b_k|^2+|b_l|^2}(b_k \ket{J,k} + b_l \ket{J,l})$, we prepared the mixed state $\rho = |b_k|^2 \ket{J,k}\bra{J,k} + |b_l|^2 \ket{J,l}\bra{J,l}$. One straightforward strategy to mitigate this is to use $\mathbb{F}$ during an initial optimization step before switching to full PI tomography, or to use full PI tomography as a verification of the parameters optimized using $\mathbb{F}$. 

~\\ \emph{Proof that $\rho_{J,m} = \sum_{i=0}^N c_{i,m} J_z^i$}. First, note that in the basis of the Dicke states $\{ \ket{J=N/2,i = -J \ldots J} \}$, $\rho_{J,m}$ is always a diagonal matrix with a one at position $(m+N/2+1,m+N/2+1)$. Because $J_z$ is diagonal, ${\rm diag}(J_z) = [-N/2,-N/2+1,\dots,N/2-1,N/2]$ and its powers are simply ${\rm diag}(J_z^i) = [(-N/2)^i,(-N/2+1)^i,\dots,(N/2-1)^i,(N/2)^i]$, with every $J_z^i$ for $0 \leq i \leq N$ linearly independent. The equation $\rho_{J,m} = \sum_{i=0}^N c_{i,m} J_z^i$ can be vectorized into a set of $N+1$ linear equations with $N+1$ variables by defining $x = \{c_{i,m}\}_{i=0}^N$, $b = {\rm diag}(\rho_{J,m})$ and the $(N+1)\times(N+1)$ coefficient matrix $A = \{{\rm diag}(J_z^i)\}_{i=0}^{N}$, such that the equation reduces to standard form $Ax = b$. Because the columns of $A$ are each from a given $J_z^i$ for $0 \leq i \leq N$, all columns are linearly independent. Thus $A$ is invertible, yielding the unique solution $x = A^{-1}b$. 

\else 
\nocite{bondEffectMicromotionLocal2022a,tothPracticalMethodsWitnessing2009,PhysRevLett.127.010504,PhysRevLett.85.194,tothPermutationallyInvariantQuantum2010,marciniakOptimalMetrologyProgrammable2022a,mccleanTheoryVariationalHybrid2016,peruzzoVariationalEigenvalueSolver2014}

\fi 

\end{document}